\documentclass[twocolumn,preprintnumbers,amsmath,amssymb,showpacs]{revtex4}

\usepackage{graphicx}
\usepackage{dcolumn}
\usepackage{bm}
\usepackage{tikz}
\usepackage{pgfplots}

\newcommand{\e}{{\rm e}}

\begin{document}


\title{Giant component sizes in scale-free networks with power-law degrees and cutoffs}

\author{A.J.E.M. Janssen}
\author{J.S.H. van Leeuwaarden}%
\affiliation{%
Eindhoven University of Technology
}%


\date{\today}

\begin{abstract}
Scale-free networks arise from power-law degree distributions. Due to the finite size of real-world networks, the power law inevitably has a cutoff at some maximum degree $\Delta$.  We investigate the relative size of the giant component $S$ in the large-network limit.
We show that $S$ as a function of $\Delta$ increases fast when $\Delta$ is just large enough for the giant component to exist, but increases ever more slowly when $\Delta$ increases further. This makes that while the degree distribution converges to a pure power law when $\Delta\to\infty$, $S$ approaches its limiting value at a slow pace. The convergence rate also depends on the power-law exponent $\tau$ of the degree distribution.
The worst rate of convergence is found to be for the  case $\tau\approx2$, which concerns many of the real-world networks reported in the literature.
\end{abstract}

\maketitle

\section{Introduction}
Many complex systems can be modeled as networks of nodes joined in pairs by edges.
Examples include the internet and the worldwide web, biological networks, the brain, neural networks, communication
and transport networks, and social networks. The relevance of networked systems
has spurred a considerable interest in developing mathematical models that can capture universal network properties.
Statistical analysis of network data suggests that many networks possess a power-law degree distribution \cite{Clauset2009,Newm10a,Pastor2002,email2002}, so that the probability $p_k$ that a node has $k$ neighbors scales with $k$ as $p_k\sim c k^{-\tau}$ for some constant $c$ and characteristic exponent $\tau>0$. The power-law distribution leads to scale-free behavior such as short distances due to the likely presence of {\it hubs} or high-degree nodes.

Many scale-free networks are reported to have an exponent $\tau$ between 2 and 3 \cite{AlbJeoBar99a,FalFalFal99,Jeoetal00}, so that the second moment of the degree distribution diverges in the infinite-size network. The existence of hubs make these networks ultra-small: The hubs each link to a large number of small-degree nodes, creating short distances and giant components of connected nodes.



Although scale-free networks have an underlying degree distribution that follows a
power law, some imposed constraint will place
a limit on the maximum degree $\Delta$, causing the power-law degree distribution to end by a rapid drop at some large degree.
This maximum degree $\Delta$ is called the {\it cutoff}. The cutoff can be seen as an inherent property of a physical system, for instance due to time constraints (the finite lifetime of a scientist puts a cap on the number of collaborations \cite{Newm01a}), or as an increasing function of the window size in the case of growing networks such as the internet or social networks \cite{barabasireview,nsw}.

In view of the inevitable presence of a cutoff, a good generic alternative for a power-law distribution is a power law with an {\it exponential}
cutoff \cite{nsw},
\begin{equation}\label{degrees}
p_k=Ck^{-\tau}\e^{-k/\Delta} \quad {\rm for}\ k\geq 1,
\end{equation}
where $\Delta$ plays the role of the maximum degree. Indeed, this distribution appears to be a good fit to data of a variety of real-world networks \cite{Pastor2002,Clauset2009},
and has been the canonical example in some of the foundational works on network theory \cite{barabasireview,nsw,Newm03a}.
 The distribution \eqref{degrees} converges to a pure power-law distribution when $\Delta\to\infty$ and to a pure exponential distribution when $\tau\downarrow 0$. Moreover, all moments are finite for any finite $\Delta$, which presents considerable technical advantages over a pure power law, because the generating functions of the degree distribution and other key measures are amenable for analysis. In contrast, for pure power laws, calculations based on the generating function formalism in \cite{nsw} diverge. But more than just a technical advantage, the cutoff $\Delta$ is one of the defining properties of networks. Even if this cutoff is only observable in the tail of the degree distribution, it can greatly influence the structural and diffusive network properties. We consider in this paper the giant component, arguably one of the most fundamental of networks properties. 


We refer with $S$ to the nondegenerate portion of network nodes contained in the giant component as the number of nodes in the network goes to infinity (large-network limit). When $\Delta=\infty$ the critical exponent $\tau_{\rm max}=3.4787\ldots$ solves $\zeta(\tau-2)=2\zeta(\tau-1)$ with $\zeta(s)=\sum_{k=1}^\infty k^{-s}$ the Riemann zeta function \cite{AieChuLu01a};  for $\tau\geq \tau_{\rm max}$, a giant component does not exist. This also means that for the distribution \eqref{degrees} with finite $\Delta$ a giant component can never exist ($S=0$) for $\tau\geq \tau_{\rm max}$. However, for all values of $\tau<\tau_{\rm max}$, which covers virtually all networks reported in the literature, there will exist a critical cutoff value $\Delta_c(\tau)$ from which point onwards a giant component exists. We henceforth assume $\Delta>\Delta_c(\tau)$ and consider the fraction $S$ of the network filled by the giant component. It is well known that $S$ strongly depends on $\tau$. In the absence of a cutoff ($\Delta=\infty$) $S$ equals 1 for $\tau\in(0,2]$ and reaches a non-degenerate limit in $(0,1)$ for $\tau\in(2,\tau_{\rm max})$. But this changes drastically for finite $\Delta$, when it is the delicate interplay between $\tau$ and $\Delta$ that determines the properties of the giant component in the large-network limit.
When $\Delta$ passes $\Delta_c(\tau)$ and the giant component exists (cf.~Figure \ref{fig22}), the giant component initially grows quickly as a function of $\Delta$. In the later stages, when $\Delta$ increases even further, the giant component will converge to its limit as $\Delta\to\infty$, but then the convergence to the limiting value becomes slow. We derive several analytic results for the rate of convergence as a function of both $\tau$ and $\Delta$. A remarkable fact revealed by our analysis is that the worst rate of convergence is obtained for $\tau=2$, when the difference between $S$ and its limit $1$ only decreases at rate $1/\ln \Delta$. This results contrasts with the classical interpretation of $\tau=2$ as the dividing line between two fundamentally different worlds of networks: For $\tau<2$ there is only one hub possible that dominates the entire network (without cutoff) leading to anomalous model behavior \cite{delgenio,Hofs15}, while for $\tau>2$ the scale-free behavior is aligned with the random graph model and its associated network properties. But the slow convergence property says that $\tau=1.5$ and $\tau=2.5$ converge at a comparable speed, while $\tau\approx 2$ converges extremely slowly. The introduction of a cutoff $\Delta$ thus changes the two-worlds picture and allows us to characterize the rapid launch and slow converge properties for all $\tau>0$. The cutoff acts as a {\it critical window} that can be used to derive the network properties, even for the less understood and often intractable critical cases $\tau=2$ and $\tau=3$.

The remainder of this paper is structured as follows. In Section \ref{sec:ran} we present the random graph model and an implicit function that describes the giant component. In Section \ref{sec:rou} we use this implicit function to establish the rate of convergence of the giant component as a function of the cutoff. We show that there are two different regimes, depending on whether the second moment of the degree distribution is finite or infinite. In Section \ref{sec:pre} we specialize to power-law degrees and use the implicit function to obtain more precise convergence rates. These analytical results show that the convergence can be extremely slow, which is also confirmed by simulations. In Section \ref{sec:con} we conclude and relate our findings to some earlier works on random graphs with cutoffs.

\section{Random graphs with cutoff: model and method}\label{sec:ran}
We consider the configuration model with degree distribution
\begin{equation}\label{degrees2}
p_k= c\tilde  p_k \e^{-k/\Delta}, \quad k\geq 0
\end{equation}
with $\tilde p_1>0$ and $\tilde p_k$ such that $p_k\to 0$ for all finite $\Delta$.
The constant $c$ is fixed by normalization and \eqref{degrees} is the special case $\tilde p_k=k^{-\tau}$ for $k\geq 1$.
A network of $N$ nodes is then constructed by first generating an i.i.d.~sequence of $N$ degrees from the distribution \eqref{degrees2}.
Half-edges or stubs are attached to each of the nodes according to this sequence, and pairs of randomly chosen stubs are turned into edges. In this way, the configuration model creates a randomly sampled network out of all possible networks with a given degree distribution.

Most analytical results for the configuration model are obtained under the assumption of locally tree-like approximations, ignoring the presence of double edges and cycles. We also use this {\it tree ansatz} and the generating function formalism to obtain analytical results, and provide additional justification of our findings through extensive simulations. Let $P(x)=\sum_{k=0}^\infty \tilde p_k x^k$ and $a=\e^{-1/\Delta}$. Under the tree ansatz, the giant component can be expressed in terms of the generating functions $G_0(x)$ and $G_1(x)=G_0'(x)/G_0'(1)$ with
\begin{equation}\label{twww}
G_0(x)=\sum_{k=0}^\infty p_k x^k = \frac{P(x a)}{P(a)}.
\end{equation}
For $G'_1(1)>1$ or equivalently $\sum_k k(k-2)p_k>0$, there exists a giant component \cite{MolRee95,nsw}. Since $\sum_k k(k-2)p_k$ increases in $\Delta$, there is thus a critical value $\Delta_c$ as of which there is a giant component. The criticality assumption $G'_1(1)>1$ is equivalent with the assumption that $a=\e^{-1/\Delta}>\xi$ with $\xi\in(0,1)$ the point at which the function $u/G_1(u)$ attains its unique maximum. Notice that $G_1(\xi)=\xi G_1'(\xi)$ and $\xi=\e^{-1/\Delta_c}$. Now assume $\Delta>\Delta_c$ or $a>\xi$ and recall that $S$ denotes the fraction of the graph occupied by the giant component. Then \cite{nsw}
\begin{equation}\label{sss}
S=1-G_0(u),
\end{equation}
where $u=u(a)\in(0,1)$ is the unique smallest nonnegative real solution of
\begin{equation}\label{uuu}
u=G_1(u).
\end{equation}

A short computation, using the definition of $G_1(x)$, shows that
we can write \eqref{uuu} as $\varphi(ua)=\varphi(a)$ with
$\varphi(x)=x/P'(x)$. The function $\varphi(x)$ is unimodal, increasing in $x\in(0,\xi)$ and decreasing in $x\in(\xi,1)$. For any $a\in (\xi,1)$ there is a unique solution $u=u(a)\in(0,1)$ of the equation $\varphi(ua)=\varphi(a)$, viz.
\begin{equation}\label{uuu2}
u=\frac{1}{a}\varphi^{-1}(\varphi(a)),
\end{equation}
where $\varphi^{-1}(\cdot)$ is the inverse function of $\varphi(x)$, $0\leq x\leq \xi$.
Observe that the function $\varphi$ that needs to be inverted does not depend on $a$; only its argument that is to be used in \eqref{uuu2} depends on $a$.
We will leverage the implicit representation of $u$ in \eqref{uuu2} to study the growth of the giant component as function of $a$.


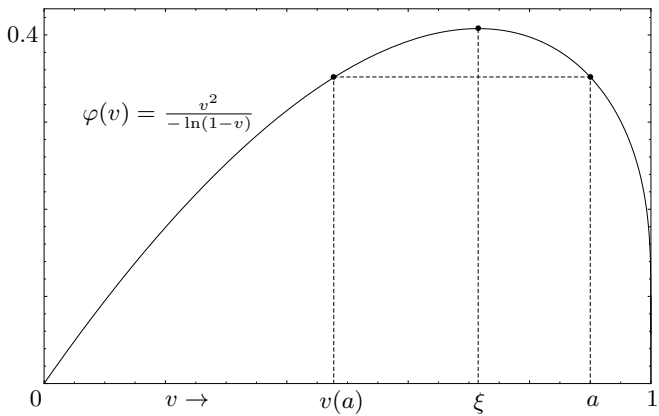
\begin{figure}[!hbtp]
\begin{center}
\begin{tikzpicture}
\node[anchor=south west,inner sep=0] at (0,0) {
\includegraphics[width=1.0\linewidth, keepaspectratio]{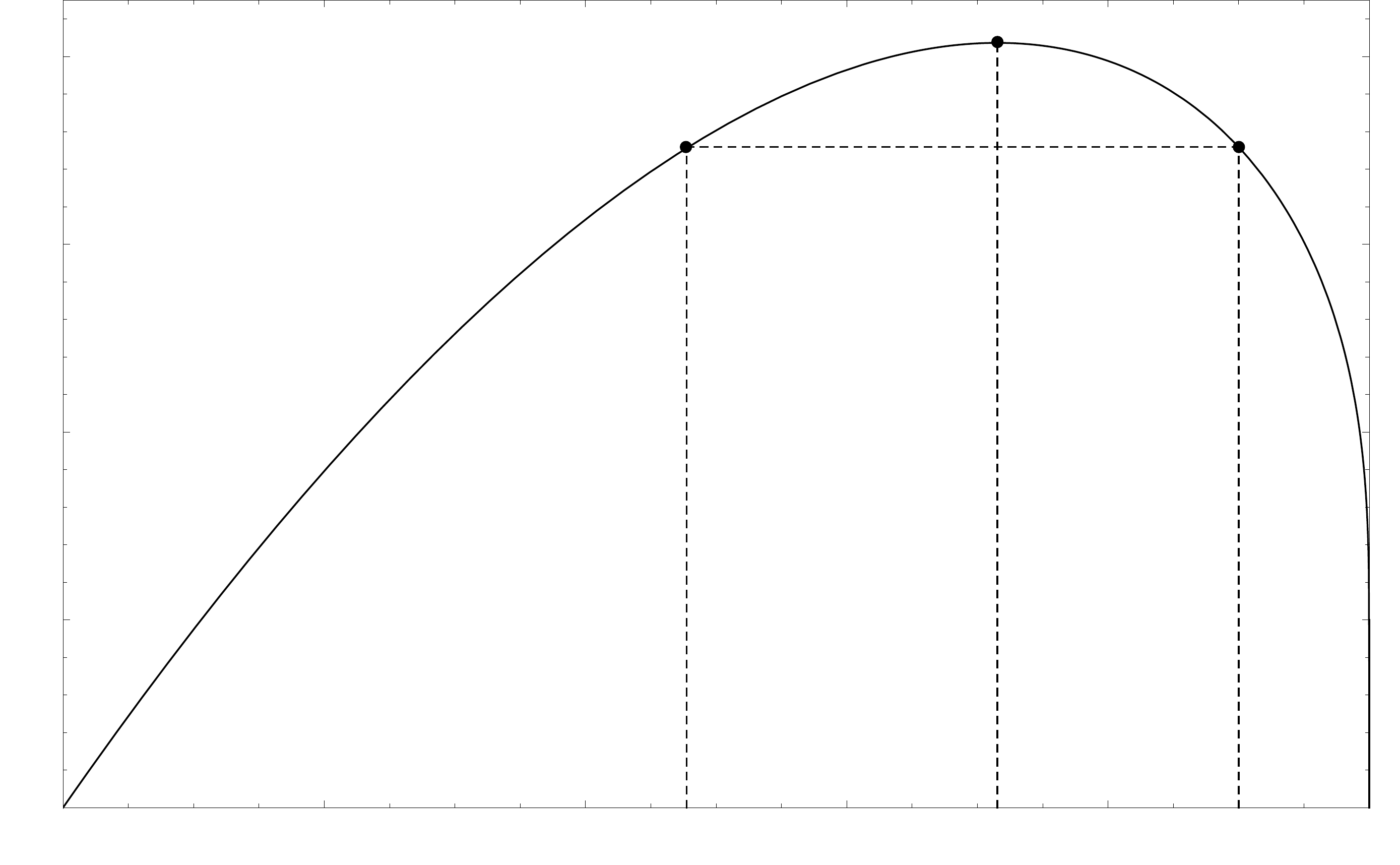}};
\node[anchor=south west,inner sep=0] at (0.2,0.05) {$0$};
\node[anchor=south west,inner sep=0] at (8.4,0.05) {$1$};
\node[anchor=south west,inner sep=0] at (-0.1,4.9) {$0.4$};
\node[anchor=south west,inner sep=0] at (2,0.05) {$v\rightarrow$};
\node[anchor=south west,inner sep=0] at (4.05,-0.05) {$v(a)$};
\node[anchor=south west,inner sep=0] at (6.10,-0.05) {$\xi$};
\node[anchor=south west,inner sep=0] at (7.6,0.05) {$a$};
\node[anchor=south west,inner sep=0] at (0.9,3.7) {$\varphi(v)=\frac{v^2}{-\ln(1-v)}$};
\end{tikzpicture}
\caption{The function $\varphi(v)$ for $\tau=2$}
\label{fig11}
\end{center}
\end{figure}

%
%
%

From \eqref{twww}, \eqref{sss} and \eqref{uuu2} we see that
\begin{equation}\label{3111}
S=1-\frac{P(v(a))}{P(a)},
\end{equation}
where
$v(a)=\varphi^{-1}(\varphi(a))$ when $a>\xi$, in which case $v(a)<\xi<a<1$ (see Figure \ref{fig11}). We can find an expression for $v=v(a)$ by using Lagrange inversion to solve the equation $\varphi(v)=y$ with $y=\varphi(a)$, which gives for $|y|<\varphi(\xi)$
the infinite series expression \cite{nsw}
\begin{equation}\label{blS}
S=1-G_0(0)-\frac{1}{P(a)}\sum_{s=1}^\infty\frac{\varphi(a)^s}{s!}\Big(\frac{{\rm d}^{s-1}}{{\rm d}x^{s-1}}\Big)\Big[(P'(x))^{s+1}\Big]_{x=0}.
\end{equation}
Although interesting, the series \eqref{blS} does not shed much light on the effect of $a$ and merely serves a numerical purpose.

 \section{A dichotomy in convergence rates}\label{sec:rou}
We now use \eqref{uuu2} to obtain information on the rate at which the giant component reaches its  limit as $\Delta\to\infty$, i.e.~$a\to 1$.
Consider the case that $P'(1)=\infty$, which applies to a degree sequence $\{\tilde p_k\}$ with a divergent first moment.  If $P'(a)\to\infty$ as $a\to 1$ then $v(a)\downarrow 0$ as $a\uparrow 1$. Since
\begin{equation}
\varphi(v)=\frac{v}{\tilde p_1+2 \tilde p_2 v+\ldots}=\frac{v}{\tilde p_1}+O(v^2), \quad v\downarrow 0,
\end{equation}
and $\varphi^{-1}(w)=\tilde p_1 w + O(w^2)$ as $w\downarrow 0$, we get
\begin{equation}
v(a)=\tilde p_1 \varphi(a)+O(\varphi^2(a))=\tilde p_1 \frac{a}{P'(a)}+O\Big(\frac{1}{(P'(a))^2}\Big).
\end{equation}
Using
\begin{equation}
\frac{P(v(a))}{P(a)}=\frac{\tilde p_1 v(a)+\tilde p_2 (v(a))^2+\ldots}{P(a)}=\frac{\tilde p_1^2}{P(a)P'(a)}+\ldots
\end{equation}
gives 
%
 \begin{equation}\label{rate1}
S=1-\frac{\tilde p_1^2}{P(a)P'(a)}-\ldots=1-O\Big(\frac{1}{P(a)P'(a)}\Big).
\end{equation}

Next consider the case $P'(a)\to P'(1)<\infty$ as $a\to 1$, so that the first moment of the degree distribution is finite.
   In that case $v(a)\downarrow \varphi^{-1}(1/P'(1))=v(1)\neq 0$ as $a\uparrow 1$. Notice that
\begin{align}\label{above}
v(a)&= \varphi^{-1}\Big(\frac{1}{P'(1)}+\Big(\frac{a}{P'(a)}-\frac{1}{P'(1)}\Big)\Big)\nonumber \\
 &=v(1)+\Big(\frac{a}{P'(a)}-\frac{1}{P'(1)}\Big)\left(\varphi^{-1}\right)'\Big(\frac{1}{P'(1)}\Big)+\ldots\nonumber \\
  &=v(1)+\Big(\frac{a}{P'(a)}-\frac{1}{P'(1)}\Big)\frac{1}{\varphi'(v(1))}+\ldots
\end{align}
Then
\begin{align}
\frac{P(v(a))}{P(a)}&=\frac{P(v(1)+(v(a)-v(1))}{P(1+(a-1))}\nonumber\\
&=\frac{P(v(1))+(v(a)-v(1))P'(v(1))+\ldots}{P(1)+(a-1)P'(1)+\ldots}\nonumber\\
&=\frac{P(v(1))}{P(1)}\Big(1+\frac{(v(a)-v(1))P'(v(1))}{P(v(1))}\nonumber\\
&-(a-1)\frac{P'(1)}{P(1)}+\ldots\Big)
\end{align}
in which we enter
$$v(a)-v(1)=(\frac{a}{P'(a)}-\frac{1}{P'(1)})\frac{1}{\varphi'(v(1))}+\ldots$$ from \eqref{above} to get
\begin{equation}\label{rate2}
S=1-\frac{P(v(1))}{P(1)}+O\Big((1-a)\frac{P''(a)}{(P'(a))^2}\Big).
\end{equation}

The big-$O$ terms in \eqref{rate1} and \eqref{rate2} thus quantify the convergence rates, which turn out to be different depending on whether the first moment diverges or not.

\section{Extremely slow convergence for power-law degrees}\label{sec:pre}

To investigate the convergence rate further, we next specialize to
 the case of power-law degrees with cutoff in  \eqref{degrees} and note that  \eqref{rate1} applies to $\tau\leq 2$ and \eqref{rate2} to $\tau>2$. For all $\tau$ we then want to characterize the convergence of $S$, as a function of $\tau$ and the cutoff $a=\e^{-1/\Delta}$.

Recall that  $\tau_{\rm max}=3.4787\ldots$ uniquely solves $\zeta(\tau-2)=2\zeta(\tau-1)$. Assuming $\tau<\tau_{\rm max}$, the point that marks the phase transition rendered by $G'_1(1) = 1$ can again be rephrased in $a=\xi$ with $\xi=\xi(\tau)\in (0,1)$ the unique solution of the equation
\begin{equation}
{\rm Li}_{\tau-2} (\xi)=2{\rm Li}_{\tau-1}(\xi)
\end{equation}
with ${\rm Li}_s(y)=\sum_{k=1}^\infty k^{-s}y^k$ the polylogarithm of order $s$ evaluated at $y$.
From \eqref{3111} we get
\begin{equation}\label{lc}
S=1-\frac{{\rm Li}_{\tau} (v(a))}{{\rm Li}_{\tau}(a)}
\end{equation}
with $v(a)$ the solution $v<\xi$ of $\varphi(v)=\varphi(a)$, where $\varphi(x)$ in the present case reads
\begin{equation}\label{4333}
\varphi(x)=\frac{x}{{\rm Li}_{\tau}'(x)}=\frac{x^2}{{\rm Li}_{\tau-1}(x)}.
\end{equation}
For all $\tau\leq 2$ we have ${\rm Li}_{\tau-1}(a)\to\infty$ as $a\to 1$, and this gives $v(a)\to 0$ and hence $S\to 1$.
When $2<\tau<\tau_{\rm max}$ we have that  ${\rm Li}_{\tau-1}(a)\to\zeta(\tau-1)<\infty$ as $a\to 1$, and then $S$ increases from 0 at $a=\xi$ to
\begin{equation}\label{4444}
S_\infty=1-\frac{{\rm Li}_{\tau} (v_\infty)}{\zeta(\tau-1)}; \quad v_\infty=\varphi^{-1}\left(1/\zeta(\tau-1)\right).
\end{equation}


When $\tau\in(0,2)$ or $(2,\tau_{\rm max})$, the speed by which $S$ reaches its limit is equal to the speed by which $1/{\rm Li}_{\tau-1}(a)$ reaches its limit as $a\to 1$. For $\tau=2$ we have that ${\rm Li}_{\tau-1}(a)=-\ln (1-a)$, and so $S$ reaches its limit $1$ only at rate $1/\ln \Delta$.

We now conduct an analysis of ${\rm Li}_{\tau-1}(a)$ when $a\to 1$ and $\tau$ is near $2$. It is here that we invoke Bateman's result \cite[\S1.11(8)]{bateman} for the polylogarithm that says for $|\ln y|<2\pi$ and $s\neq 1,2,\ldots$
\begin{equation}\label{bate}
{\rm Li}_s(y)=\Gamma(1-s)\Big(\ln y^{-1}\Big)^{s-1}+\sum_{r=0}^{\infty}\zeta(s-r)\frac{(\ln y)^r}{r!}.
\end{equation}
When $s=1,2,\ldots$ the term in the series \eqref{bate} with $r=s-1$ and the term involving $\Gamma(1-s)$ must be combined (see below).
Observe that while the series ${\rm Li}_{\tau-1}(a)=\sum_{k=1}^\infty k^{-\tau+1}a^k$
 converge ever more slowly as $a$ increases towards 1, Bateman's formula \eqref{bate} transforms this slowly converging series into an expansion with only a few dominant terms when $a\to 1$. We obtain for instance from \eqref{bate} that
\begin{equation}\label{form2}
{\rm Li}_{\tau-1}(a)=\Gamma(2-\tau)\Delta^{2-\tau}+\zeta(\tau-1)+O(\Delta^{-1}),
\end{equation}
so that
\begin{equation}\label{form5}
\frac{1}{{\rm Li}_{\tau-1}(a)}=M(\tau)+O(\Delta^{-|\tau-2|}), \quad \Delta\to\infty
\end{equation}
with $M(\tau)=1/\zeta(\tau-1)$ for $\tau>2$ and $M(\tau)=0$ for $\tau<2$.
The formula \eqref{form5} shows that the limit $M(\tau)$ is reached more slowly as $\tau$ gets closer to $2$, either from above or below.

Near $\tau=2$, both $\Gamma(2-\tau)$ and $\zeta(\tau-1)$ in \eqref{form2} are large but have opposite sign. This has a further detrimental effect on the convergence speed when $\tau$ is near $2$. More precisely,
\begin{equation}\label{form7}
\Gamma(2-\tau)=\frac{1}{2-\tau}-\gamma+O(2-\tau)
\end{equation}
and
\begin{equation}\label{form8}
\zeta(\tau-1)=\frac{1}{\tau-2}+\gamma+O(2-\tau)
\end{equation}
with $\gamma=0.5772\ldots$ Euler's gamma. Notice that \eqref{form8} implies that
$M(\tau)=\tau-2+O\left((\tau-2)^2\right)$, $\quad \tau\downarrow 2$,
and so the limit function is continuous at $\tau=2$. Next, from \eqref{form2}, \eqref{form7} and \eqref{form8} we see that
\begin{align}\label{form10}
{\rm Li}_{\tau-1}(a)&=\Big(\frac{1}{2-\tau}-\gamma\Big)\left(\Delta^{2-\tau}-1\right)\nonumber\\
&+O\left((2-\tau)\left(
\Delta^{2-\tau}+1\right)\right)+O(\Delta^{-1}).
\end{align}
 The leading term in the approximation \eqref{form10}
contains the salient features of ${\rm Li}_{\tau-1}(a)$ as $a\to 1$ and $\tau$ near 2.
For fixed $\Delta\geq 1$,
\begin{equation}\label{}
\lim_{\tau\to 2}\Big(\frac{1}{2-\tau}-\gamma\Big)\left(\Delta^{2-\tau}-1\right)
=\ln \Delta,
\end{equation}
irrespective whether $\tau$ approaches $2$ from above or below.
 Note that ${\rm Li}_{1}(a)=\ln\Delta+O(\Delta^{-1})$. Furthermore, for fixed $\tau<2$ and $\Delta\to\infty$,
 \begin{equation}\label{form13}
\Big(\frac{1}{2-\tau}-\gamma\Big)\left(\Delta^{2-\tau}-1\right)
=\frac{\Delta^{2-\tau}}{2-\tau}+O(1),
\end{equation}
while for $\tau>2$
 \begin{align}\label{form14}
\lim_{\Delta\to\infty}\Big(\frac{1}{2-\tau}&-\gamma\Big)\left(\Delta^{2-\tau}-1\right)
=\frac{1}{\tau-2}+\gamma\nonumber\\&=\zeta(\tau-1)+O(\tau-1).
\end{align}

In its most rudimentary form we obtain from \eqref{form10} the approximation
\begin{equation}\label{form15}
\frac{1}{{\rm Li}_{\tau-1}(a)}=\frac{2-\tau}{\Delta^{2-\tau}-1}
\end{equation}
with a relatively small error, and this describes accurately the relaxation rate of $S$ when passing from the range $\tau<2$ to $\tau>2$. At the point $\tau=2$ the approximation \eqref{form15} is to be interpreted as its limiting value $1/\ln\Delta$, and this confirms the extremely slow convergence of $S$ in this case. For comparison, for $\tau$ near $1$ \eqref{bate} gives \begin{equation}\label{}
{\rm Li}_{\tau-1}(a)=\Delta^{2-\tau}+O(1)+O((1-\tau)\Delta^{2-\tau}),
\end{equation}
so that $1/{{\rm Li}_{\tau-1}(a)}\approx \Delta^{\tau-2}$, while for $\tau$ near $3$ we get
\begin{equation}\label{}
{\rm Li}_{\tau-1}(a)=\zeta(2)-\frac{1}{\Delta}\frac{\Delta^{3-\tau}-1}{3-\tau}+O(\tau-3)+O(\Delta^{2-\tau}),
\end{equation}
and hence
\begin{equation}\label{}
\frac{1}{{\rm Li}_{\tau-1}(a)}\approx \frac{1}{\zeta(2)}\Big(1+\frac{1}{\Delta}\frac{\Delta^{3-\tau}-1}{3-\tau}\Big)\stackrel{\tau=3}{=}\frac{1}{\zeta(2)}\Big(1+\frac{\ln \Delta}{\Delta}\Big).
\end{equation}
For both $\tau=1$ and $\tau=3$ convergence is much faster than $1/\ln\Delta$.

\begin{figure}[!hbtp]
\begin{center}
\begin{tikzpicture}
\node[anchor=south west,inner sep=0] at (0,0) {
\includegraphics[width=1.0\linewidth, keepaspectratio]{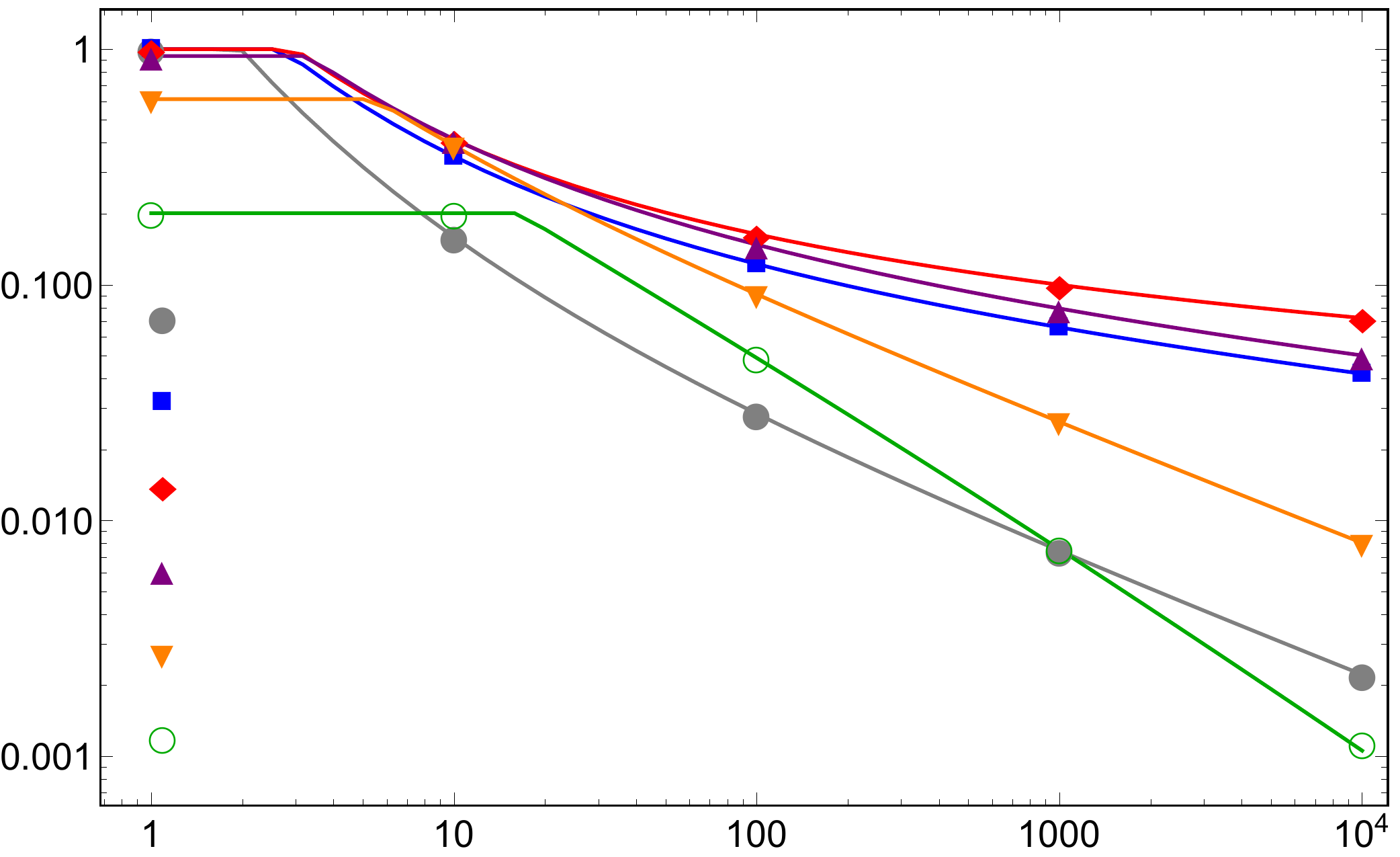}};
\node[anchor=south west,inner sep=0] at (3.5,-0.2) {$\Delta\rightarrow$};
\node[anchor=south west,inner sep=0] at (7,4.6) {$S_\infty-S$};
\node[anchor=south west,inner sep=0] at (1.2,3.21) {$\tau=1.5$};
\node[anchor=south west,inner sep=0] at (1.2,2.70) {$\tau=1.9$};
\node[anchor=south west,inner sep=0] at (1.2,2.17) {$\tau=2$};
\node[anchor=south west,inner sep=0] at (1.2,1.66) {$\tau=2.1$};
\node[anchor=south west,inner sep=0] at (1.2,1.15) {$\tau=2.5$};
\node[anchor=south west,inner sep=0] at (1.2,0.64) {$\tau=3$};
\end{tikzpicture}
\caption{The difference $S_\infty-S$ as measured in numerical simulations
for different values of
 $\tau$ and $\Delta$. Each point is an average over $10^4$ different
network realizations of size $N=10^6$. Solid lines are the approximate analytic solutions
given by \eqref{sss}. 
}
\label{fig22}
\end{center}
\end{figure}
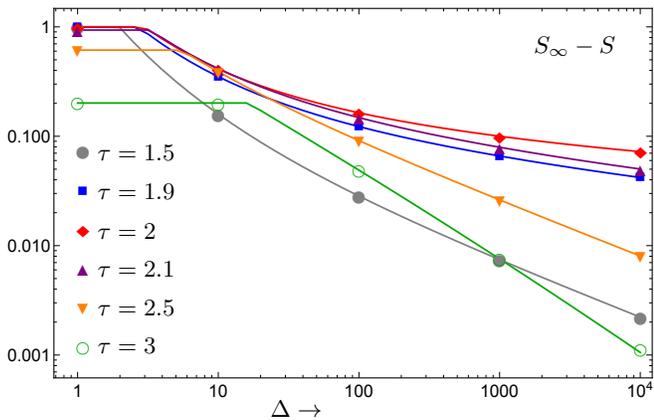

When $\tau$ is sufficiently away from $2$ we see from \eqref{form15} that there is the relation
\begin{equation}\label{eqlin}
\ln(S_\infty-S)=C-|\tau-2|\ln\Delta
\end{equation}
with $C$ a constant that depends on $\tau$ and can be determined using \eqref{rate1} and \eqref{rate2} and where $S_\infty$ is to be interpreted as $1$ for $\tau\leq 2$.
Figure \ref{fig22} shows a comparison between extensive simulations and the analytical solution in \eqref{sss} for $\ln (S_\infty-S)$ as a function of $\ln \Delta$.
For $\tau$ sufficiently away from $2$ the linear rate as predicted in \eqref{eqlin} is confirmed. Also, when $\tau$ approaches $2$, the pattern corresponds to the $1/\ln \Delta$ convergence rate. This would be even better visible when plotting $\ln(S_\infty-S)$ against $\ln\ln\Delta$. In this case only $\tau=2$ shows a linear relation against $\ln\ln \Delta$, stressing the fact that the $S$ reaches its limit extremely slowly. This implies that in practice, for scale-free networks with $\tau\approx 2$, the effect of a cutoff cannot be neglected, even for large cutoffs. Observe also that, as predicted from our analysis, it is the deviation from $2$ rather than the absolute value of $\tau$ that determines the convergence rate.

\section{Conclusion}\label{sec:con}

To summarize, we studied in this paper the fundamental role of cutoffs in scale-free networks and considered the giant component as a function of the cutoff $\Delta$. An important technical step was to convert series that were slowly converging in $\Delta$ into expansions of which the leading terms are dominant for large $\Delta$. 
This led to a characterization of the rate at which the giant component reaches its limit as the cutoff increases. We found that the rate of convergence is overall slow, and the worst for power-law degree distributions with $\tau\approx 2$.

Let us relate these insights to some widely cited real-world networks. The famous example of the e-mail network studied in \cite{email2002}, which connects two people if they were either source or destination of an e-mail, gave rise to $\tau=1.81$, $\Delta\approx 100$, $59.912$ nodes and $S=0.95$. For $\tau=1.81$ slow convergence is expected, and while the cutoff is already 100, it should be increased quite substantially before the giant component will cover the final 5\% of the network. Collaboration networks defined as graphs with an edge connecting two people if they have written a paper together were studied in \cite{Newm01a}. Two example data sets in \cite{Newm01a} concerned publications in astrophysics (among 16,706 authors) and condensed matter physics (among 16,726 authors) to which power laws with exponential cutoffs were fitted with the triplet $(\tau,\Delta,S)$ equal to $(0.91,49,0.89)$ for astrophysics and $(1.1,15.7,0.85)$ for condensed matter physics. With these extreme exponents of $0.91$ and $1.1$ it is perhaps surprising that the giant components contain only 89\% and 85\% of the networks, but in view of the slow convergence reported in this paper and the moderate cutoff values, this can be understood. Another illustrious example concerns the World Wide Web consisting of web pages and links between them. The WWW was found to have power-law indegrees with $\tau=2.1$ and $\Delta=900$ \cite{AlbJeoBar99a}. A similar exponent was reported for the Internet, consisting of physically connected routers, and in \cite{FalFalFal99} found to have a $\tau$ in the range 2.1-2.2 and $\Delta$ in the range 30-40. Metabolic E. coli networks built up of substrates that are connected to one another through links, which are the actual metabolic reactions, were again found to have a similar exponent $\tau=2.2$ and this time cutoff $\Delta=110$.

Apart from the many networks for which cutoffs have been shown to be statistically relevant \cite{barabasireview,nsw,email2002}, it seems likely that in many cases where pure power-law distributions have been observed like in \cite{AlbJeoBar99a,FalFalFal99,Jeoetal00} with exponents $\tau$ near $2$,  even better models would be obtained with an exponential cutoff with $\Delta$ some large fixed value. Unlike models with a conjectured pure power-law distribution, models with a cutoff can always capture the inherent maximum degree of a networks, and then by letting this maximal degree tend to infinity it can be explored to what extend the cutoff, say very large yet finite, changes the fundamental properties of complex networks. Ignoring the cutoff leaves unexplained network phenomena that arise due to the limited size of networks, and hence the finite-size effects of the cutoff.

{\it Acknowledgment.} We thank Marko Boon for performing the extensive simulations leading to Figure \ref{fig22}. We thank Remco van der Hofstad for proofreading and providing several useful comments. This work was financially supported by The Netherlands Organisation for Scientific Research (NWO) through a TOP grant and the Gravitation Project NETWORKS and by The European Research Council (ERC) through a Starting Grant.

\bibliography{bibo}

\end{document}